%% file: main.tex
\documentclass[sigconf]{acmart}
\usepackage{url}
\usepackage{todonotes}
\usepackage{float}
\usepackage{multicol}
\usepackage{tikz} 
\usepackage{pgfplots}
\usepackage{enumerate}
\pgfplotsset{compat=newest}
\usepackage{amssymb}
\usepackage{booktabs}
\usepackage{csquotes}
\usepackage{footnote}
\makesavenoteenv{tabular}
\makesavenoteenv{table}
\newenvironment{widequotation}{\list{}{\listparindent 1.5em \itemindent\listparindent
		\rightmargin 0pt \parsep 0pt plus 1pt}\item\relax}
{\endlist}

\usepackage{xspace}
\def\signed#1{{\leavevmode\unskip\nobreak\hfil\penalty50\hskip2em
		\hbox{}\nobreak\hfil\raise-3pt\hbox{#1}
		\parfillskip=0pt \finalhyphendemerits=0 \endgraf}}
\newsavebox\mybox
\newenvironment{aquote}[1]
{\savebox\mybox{(#1)}\begin{widequotation}\itshape``\ignorespaces}
	{\unskip"\signed{\usebox\mybox}\end{widequotation}}

\begin{document}

\title{Security Assurance Cases for Road Vehicles: \\an Industry Perspective}

\author{Mazen Mohamad}
\affiliation{%
  \institution{Uni. of Gothenburg and Chalmers}
  \state{Sweden}
}

\author{Alexander \AA str\"om}
\affiliation{%
  \institution{Volvo Cars Corporation}
  \state{Sweden}
}

\author{\"Orjan Askerdal}
\affiliation{%
  \institution{AB Volvo}
  \state{Sweden}
}

\author{J\"orgen Borg}
\affiliation{%
  \institution{Volvo Cars Corporation}
  \state{Sweden}
}

\author{Riccardo Scandariato}
\affiliation{%
  \institution{Uni. of Gothenburg and Chalmers}
  \state{Sweden}
}

\renewcommand{\shortauthors}{Mohamad and \AA str\"om, et al.}
\renewcommand{\shorttitle}{Security Assurance Cases for Road Vehicles}

\begin{abstract}
Assurance cases are structured arguments that are commonly used  to reason about the safety of a product or service.
Currently, there is an ongoing push towards using assurance cases for also cybersecurity, especially in safety critical domains, like automotive.
While the industry is faced with the challenge of defining a sound methodology to build security assurance cases, the state of the art is rather immature.
Therefore, we have conducted a thorough investigation of the (external) constraints and (internal) needs that security assurance cases have to satisfy in the context of the automotive industry.
This has been done in the context of two large automotive companies in Sweden.
The end result is a set of recommendations that automotive companies can apply in order to define security assurance cases that are (i) aligned with the constraints imposed by the existing and upcoming standards and regulations and (ii) harmonized with the internal product development processes and organizational practices.
We expect the results to be also of interest for product companies in other safety critical domains, like healthcare, transportation, and so on.
\end{abstract}



\maketitle

\input{introduction}
\input{related_work}

\input{methodology}
\input{results-RQ1}

\input{results-RQ2}

\input{threats}
\input{discussion.tex}
\input{conclusion}

\section*{Acknowledgement}
This work is partially supported by the CASUS research project funded by VINNOVA, a Swedish funding agency.

\bibliographystyle{ACM-Reference-Format}
\bibliography{references}

\end{document}

%% file: introduction.tex
\section{Introduction}

\begin{figure}
    \centering
    \includegraphics[width=\columnwidth]{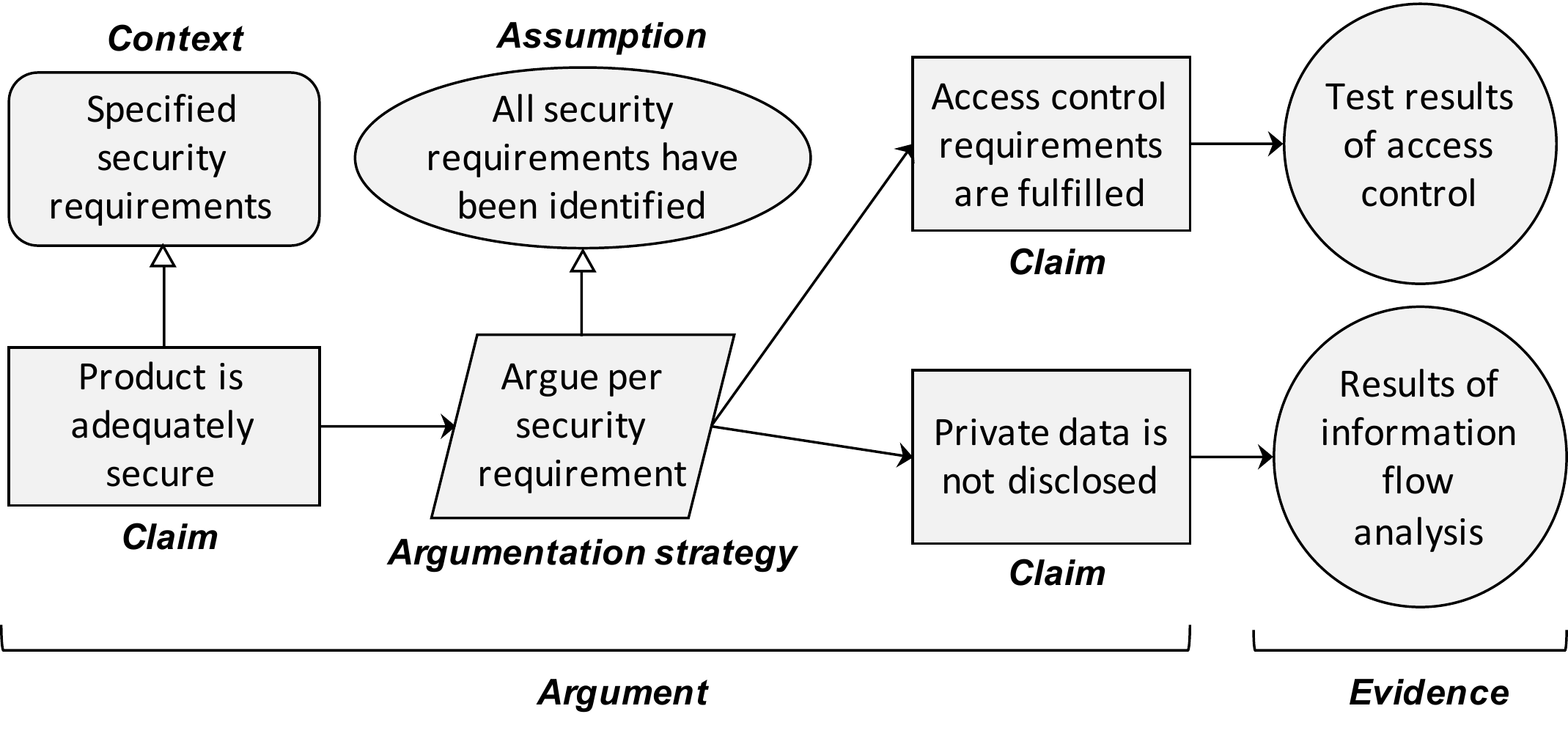}
    \caption{A simple example of a security assurance case}
    \label{fig:introSAC}
\end{figure}

An assurance case can be described as a structured set of arguments that are supported by evidence, e.g., collected from the results of the validation and verification activities \cite{goodenough2007}.
A simple example is given in Figure \ref{fig:introSAC} and the reader could recognize the resemblance with the logical argumentation of a legal case.
Assurance cases have been in use for several decades in order to argue for completeness and correctness of various dependability attributes in a wide range of industrial fields. 
In the automotive industry, assurance cases for functional safety (or safety cases) are a common practice since the release of the ISO 26262 standard on functional safety for road vehicles  in 2011 \cite{iso26262_1ed}. 
The necessity to adopt assurance cases also for cybersecurity is emerging in the automotive industry only now, especially because of the upcoming release of security standards that explicitly demand for them \cite{iso21434}.
The necessity is also felt from within the automotive industry.
The vehicle industry is going through a rapid transformation with features such as increased connectivity and automated driving as two of the major driving forces.
Features like these demands various external interfaces which exposes potential vulnerabilities in the connected devices of the vehicles and increases the risks in a way that was never seen before.
Therefore, a more systematic way to ``reason'' around security is desirable.

The push towards adopting security assurance cases represent a challenge that is both technical and organizational at the same time. 
For instance, the selection of a given argumentation strategy (i.e., the structuring of the security case) is not a purely technical choice, as it might require to re-organize the way the product is developed, e.g., in order to introduce extra activities and work products to create the necessary evidence.
Such organizational issues cannot be underestimated in large eco-systems, like the development environment in vehicle manufacturers (OEM).

In our analysis of the literature, we have found that the related work has not investigated the constraints and requirements around high-impact technical decisions such as (i) how to structure a security case, (ii) how to collaborate with suppliers on security cases, (iii) how to effectively update a security case, and so on.
Therefore, in a collaboration between one academic institution and two OEMs, we have performed a study of the industrial needs that pertain the technical choice of adopting (or defining) a methodology for security assurance cases in an automotive organization.

This paper has \emph{three main contributions}. 
First, we have performed a systematic study of the security-relevant regulations and standards in the automotive domain. 
In this analysis, we have identified the explicit and implicit constraints laid out by such documents with respect to security cases.
We call these the \emph{external forces} driving the adoption of security cases.
The analysis is performed by a pool of industrial security experts (working in two panels) who are also members of several standardization committees and, hence, know how to interpret these documents, which are, at times, somewhat fuzzy.

Second, we have performed an empirical study with a significant number of stakeholders (more than 20 people) that are affected either directly (e.g., as prospective producers) or indirectly (e.g., as prospective consumers) by security assurance cases.
By applying rigorous methods from the field qualitative research, we have systematically identified the \emph{internal organizational needs and opportunities} that must be taken in consideration for a successful adoption of a security case methodology.

Third, we have combined the observations collected in the two above-mentioned studies and translated them into a list of practical recommendations on security cases for the automotive industry.

The rest of the paper is structured as follows. 
In Section \ref{sec:rw}, we provide more background on assurance cases and discuss the related work. 
In Section \ref{sec:method}, we formulate the research questions and describe the research methodology. 
In Sections \ref{sec:rq1} and \ref{sec:rq2}, we present the results. 
In Section \ref{sec:threats}, we discuss the threats to the validity of our research.
In Section \ref{sec:discussion}, we discuss the results and provide our recommendations.
Finally, in Section \ref{sec:end}, we presents the concluding remarks.

%% file: related_work.tex
\section{Background and Related work}\label{sec:rw}

\textbf{Security Assurance Cases.}
The argumentation in a Security Assurance Case (SAC) consists of claims about security for the system in question, and the evidence to justify these security related claims.
A SAC consist of the following primary components: (i) security claims, (ii) the context in which the claims should hold, (iii) an argument about the security claim, (iv) the strategy used to build the argument, and (v) A body of evidence to prove the claims \cite{knight2015,alexander2011}
A SAC can be expressed in a textual or graphical format\cite{alexander2011}. The most common graphical formats are the Goal Structure Notation (GSN, \cite{gsn}), and the Claims Arguments, and Evidence notation (CAE, \cite{cae}). 
Researchers have been exploring several approaches for creating the argument part of SAC. Biao et al.~\cite{xu2017} suggest dividing the argument into different layers, and using different patterns to argue in each of these layers, which include an asset layer, and a threat layer. Another approach was used by Poreddy et al.~\cite{poreddy2011}, which uses different security properties as argumentation strategy. Other used approaches argue by development life-cycle phases \cite{ray2015}, standard recommendations \cite{he2012}, and product components \cite{hawkins2015}.

\textbf{Security Assurance Cases in Automotive.}
The review of literature shows very little use of assurance cases for security in the automotive industry. 
One of the very few related studies is the work by Cheah et al.~\cite{Cheah2018}. The authors present a classification of security test results using security severity ratings. These security tests then form a body of evidence used as an input for constructing a security assurance case. The study suggests a bottom-up approach for constructing a security case, but does not provide a complete example case to show how the body of evidence is connected to claims. 
Another related study is the work by Fung et al.~\cite{Fung2018} which studies maintaining assurance cases by using automated change impact analysis. A tool was created for that purpose and a case study was conducted in the automotive domain. The authors claim the tool can be used for both safety and security. However, the example case in the paper is only about safety. 


\textbf{Safety Assurance Cases in Automotive.}
Safety cases have been in use in the automotive industry for several years.
In the second edition of the ISO 26262 standard \cite{iso26262_2ed} it is stated that release for production shall only be approved if there is \emph{``sufficient evidence for confidence in the achievement of functional safety''} and that this could be provided by the safety case. This clearly shows the importance of a safety case. 
Birch et al.~\cite{Birch2013} perform an industrial case study focusing on the product-related arguments of a safety case as opposed to the process-related arguments, which according to the authors often gets the overhand. The authors discuss the outcome of the case study listing challenges and advantages. They mention how the engineers that design the system may benefit from the safety case compilation throughout the project, especially the product-related part, and address issues in a timely manner.

%% file: methodology.tex
\section{Research Methodology}\label{sec:method}

This study was conducted in two large automotive OEMs located in Sweden. \emph{Company A} is a passenger car manufacturer, while \emph{Company B} is a truck manufacturer. 

\subsection{Research Questions}

This work is motivated by the urgency that is currently perceived by the automotive industry with respect to implementing security assurance cases.
This is due to the emergence of several standards and regulations that are forcing the industry to develop a methodology for SAC in order to stay compliant and avoid legal risks.
We call these the \emph{external drivers} that will impose constraints on how SAC should look like.
Accordingly, we formulate the first research question as follows:\\
\textbf{RQ1}. What are the \textbf{constraints} for SAC coming from regulations and standards in the automotive markets of EU, US, and China?

The need to develop a strategy for SAC is also perceived by the automotive companies as an opportunity to improve their cybersecurity development process. Also, such methodology should integrate with the product lifecycle. 
As such, we have investigated these \emph{internal drivers}. Accordingly, we formulate the second research question as follows:\\
\textbf{RQ2}. What are the \textbf{needs and opportunities} related to security assurance cases in the automotive industry?

\subsection{Methodology}\label{subsec:methodology}
\begin{table*}
\caption{List of the participants}
\label{tbl:participants}
\begin{minipage}{\columnwidth*2}
\begin{center}
\begin{tabular}{llllllll}
\toprule
               &               &                              & RQ1                   & RQ2        &            &                &            \\ 
           ID  & Company       & Role                         & Analysis of standards  & Pre-study  & Workshop   & Prioritization & Interviews  \\ 
           &&&and Regulations&&&&\\
\midrule
            1  & Company A     & Attribute Leaders 
               &  \checkmark           & \checkmark &            &                &            \\
            2  & Company A     & Regulatory experts              &  \checkmark           & \checkmark &            &                &            \\
            3  & Company A     & Safety experts              &  \checkmark           & \checkmark &            &                &            \\
            4  & Company A     & Security R\&D experts              &  \checkmark           & \checkmark &            &                &            \\
            5  & Company A     & Product Owner Security              &  \checkmark           & \checkmark &            &                &            \\
            6  & Company A     & Security Engineers              &  \checkmark           & \checkmark &            &                &            \\
            7  & Company B     & Security expert              &                       &            & \checkmark & \checkmark     &            \\
            8  & Company B     & Security expert              &                       &            & \checkmark & \checkmark     &            \\
            9  & Company B     & Safety expert                &                       &            & \checkmark & \checkmark     &            \\
            10  & Company B     & Security Engineer            &                       &            & \checkmark &                &            \\
            11 & Company B     & Software architect           &                       &            & \checkmark &                &            \\ 
            12 & Company B     & Principal Engineer           &                       &            & \checkmark &                &            \\
            13 & Company B     & Software Engineer            &                       &            & \checkmark &                &            \\
            14 & Company B     & Security Engineer            &                       &            & \checkmark &                &            \\
            15 & Company B     & Security Engineer            &                       &            & \checkmark &                &            \\
            16 & Company B     & Security Engineer            &                       &            & \checkmark & \checkmark     &            \\
            17 & Company B     & Security Engineer            &                       &            & \checkmark & \checkmark     &            \\
            18 & Company B     & Security expert              &                       &            &            & \checkmark     & \checkmark (facilitator) \\
            19 & Company B     & Solution Train Engineer (STE)&                       &            &            &                & \checkmark (stakeholder) \\
            20 & Company B     & Solution Manager             &                       &            &            &                & \checkmark (stakeholder)\\
            21 & Company B     & Functional Safety Assessor   &                       &            & \checkmark &                & \checkmark (stakeholder)\\ 
            22 & Company B     & Component Owner              &                       &            &            &                & \checkmark (stakeholder)\\
            23 & Company B     & Senior Legal Counsel         &                       &            &            &                & \checkmark (stakeholder)\\
            24 & Company B     & Security expert              &                       & \checkmark &            & \checkmark     &            \\
            25 & Company B     & Security Manager             &                       & \checkmark &            & \checkmark     &            \\
            26 & Company B     & Security R\&D                &                       & \checkmark &            &                &            \\            
            27 & University C  & Researcher                   &                       &            &            & \checkmark     & \checkmark (interviewer) \\
            28 & University C  & Researcher                   &                       &            &            & \checkmark     &            \\ 
\bottomrule
\end{tabular}
\end{center}
\footnotesize
\end{minipage}
\end{table*}

As shown in Table \ref{tbl:participants}, this study involved a total of 28 participants.

\textbf{RQ1.} Concerning the analysis of the standards and regulation, \emph{Company A} maintains a knowledge base of relevant documents as part of their security governance framework.
This knowledge base consist of standards, regulations, guidelines, best practices,  etc applicable for various markets.
Furthermore, this knowledge base covers both current and upcoming trends.
We assume that such knowledge base is fairly complete, at least for the most relevant markets (e.g., US, EU, China).
Furthermore, this knowledge base includes, among other things, information regarding the categorization of  requirements, their relevance, the parts of the organization that is affected, and which life-cycle phases of the products are impacted.
This knowledge base represented the pool of documents we have analyzed in order to answer RQ1.

In particular, in this study we prioritize new regulations and standards that will soon come into effect and focus on the markets mentioned above.
The filtered documents (listed in Section \ref{sec:rq1}) have been analyzed for explicit references to security assurance cases or their parts.
We also looked for implicit relationships to SAC.
For instance, in the SELF DRIVE Act \cite{SELFDRIVE} there is a demand that manufactures must have a \textit{Cybersecurity Plan} that includes, among other things, processes for identification, assessment and mitigation of vulnerabilities (that are reasonably foreseeable). A SAC can then be used to show how this requirement is fulfilled listing the demanded processes and the evidence for them. 


\smallskip

\begin{figure}
    \centering
    \includegraphics[width=\columnwidth]{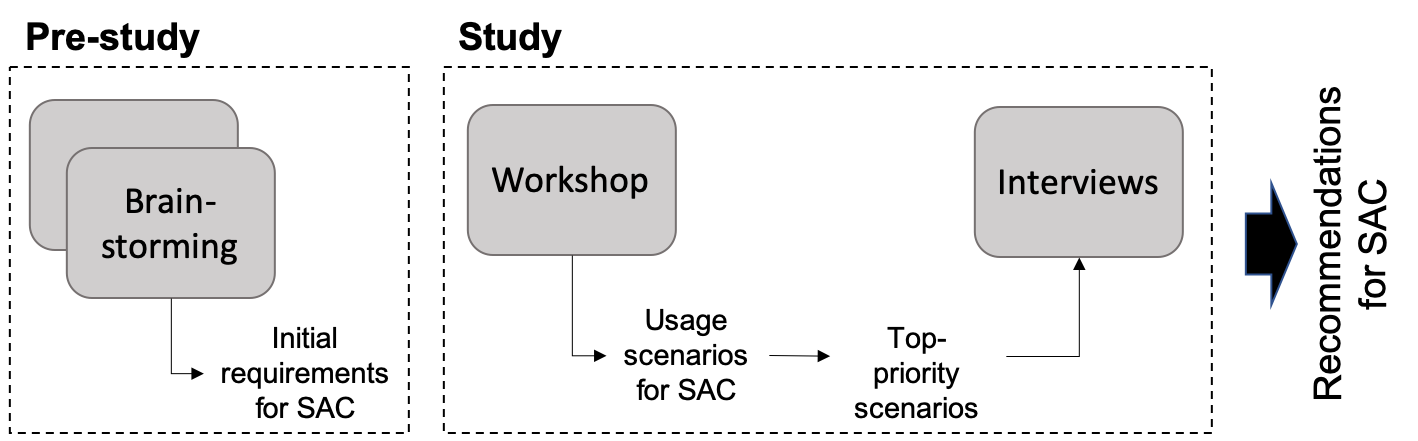}
    \caption{Method used to get the industrial needs for security assurance cases}
    \label{fig:RQ2Method}
\end{figure}

\textbf{RQ2.} To understand the internal needs for security assurance cases in the automotive industry, we used a three-steps method, as shown in Figure \ref{fig:RQ2Method}.

\paragraph{Pre-study}
The goal of the pre-study was to assess the overall industrial expectations with regards to SAC.
In particular, the pre-study reflects the point of view of the security leaders in the two companies participating to this investigation. 
In each company and independently from each other, a panel of experts performed a series of brainstorming meetings. 
The goal of the brainstorming was to form a consensual opinion of how the SAC should `look and feel', within each company and from the perspective of security people.
The results of the two panels were compared and merged into a single list of requirements and constraints, with the support of University C.

The panels consisted of R\&D personnel (i.e., technical leaders) with expertise in both security and safety.
Further, the participants were all familiar with the concept of assurance cases.
As a consequence, these groups of people were very homogeneous and might not have been aware of the full spectrum of needs and expectations in their two large companies.
Therefore, we decided to perform a larger study (comprising a workshop and a series of interviews), involving a larger and more diverse set of stakeholders\footnote{The study has been performed only in Company B, due to resource constraints.}.
In Section \ref{sec:rq2} we present the results from both the pre-study and the larger study, and we compare the observations.

\paragraph{Workshop}
The first step in our study was conducting a workshop to elicit  usage scenarios related to security assurance cases. The workshop was conducted at Company B. We invited stakeholders from different backgrounds and different parts of the organization. 
As shown in Table \ref{tbl:participants}, in total, we had 12 participants and three moderators contributing.
We started the workshop with a presentation to introduce security assurance cases to the participant which did not have previous experience with them.
We then divided the participants into three groups of 4 participants, making sure to spread similar roles and competences among the groups, e.g., we had three participant who were familiar with safety cases, so we assigned them to different groups. 
We asked the groups to brainstorm for 45 minutes on usage scenarios for security assurance cases, and to describe them as user stories, like ``As a <<role>> I would use security assurance cases for <<usage>>''~\cite{cohn2004}. We explicitly asked the participants to come up with real-life scenarios in the context of their company. 



\paragraph{Prioritization and Interviews}


At this step, we wanted to dig deeper and get a better understanding of the most important scenarios. We also wanted to acquire the point of view of more diverse stakeholders. Hence we had to prioritize the scenario and identify stakeholders to be interviewed for the top ones.

Concerning the prioritization, we aimed at getting expert opinions on which usage scenarios are of most value to the company, from a security perspective. 
As shown in Table \ref{tbl:participants}, we sent out the scenarios collected from the workshop to 10 security experts and asked them to select the top five scenarios by assigning a rank from 1 to 5 to them, where 5 is assigned to the most valuable scenario for the company.

Afterwards, we selected, the top five usage scenarios and identified a key stakeholder at company B for each.
Finally, we conducted in-person interviews with these stakeholders. Note that at the interviews, a security expert from the company was also present as a facilitator of the discussion.
The interviewees were selected based on the relevance of their expertise to the actors of the user stories in the corresponding usage scenarios. For example, the actor of one of our top usage scenarios is a \emph{legal risk owner}, hence, we selected an interviewee who has extensive experience in law and has the role: \emph{senior legal counsel} in Company B.

We organized each interview into four parts, according to the following themes:
\begin{enumerate}[i]
\item 
\textbf{value} In the first part, we focus on the value that SAC might bring to the stakeholder in terms of, e.g., efficiency, and quality management. The objective of the discussion is to picture the ‘status quo’ (e.g., to understand how the level of security is currently appraised) and the expectations (i.e., how things should improve). 
\item \textbf{content and structure} The focus of this part is to get the interviewees' technical opinions on how the content and structure of SAC should be, e.g., in terms of level of detail and types of claims; 
\item 
\textbf{integration} This part is about understanding how SAC could be integrated with the current way of working, and whether it could fit in the current activities, or would require modifications to the process; and
\item
\textbf{challenges and opportunities} The last part of the interview is about understanding the challenges and opportunities that the stakeholders foresee in applying SAC.
\end{enumerate}

In each interview, there was an interviewer (an author), an interviewee, and a security expert who acted as a discussion enabler (also an author). We recorded the interviews, and used the recordings to extract a transcript for each interview. These were then sent to the corresponding interviewees validation, and additional comments. 

%% file: results-RQ1.tex
\section{RQ1: External drivers}\label{sec:rq1}

In this section we look at external drivers that put requirements on the automotive industry. These drivers include regulations, standards, best-practises, and guidelines. The intent of the analysis is to find the relation and the motivation of a SAC from these documents. 
We look at some of the current documentation as well as some upcoming ones. We do not present a complete list, but rather look at the ones that may have a large impact on the subject and the field of interest. These are presented in~\autoref{tbl:external}. The first three columns in the table indicate how SAC is referenced in the document, i.e., whether it is explicitly or implicitly mentioned, or whether it is beneficial for the purpose of the document. The rest of the table categorizes the documents in terms of type (regulation, standard, guideline, or best-practice) and market. The last four columns indicate whether the document requires compliance or conformance, and whether the document targets the process or the product.

\begin{table*}
\caption{External drivers and references}
\label{tbl:external}
\begin{tabular}{llll|llllll}
\toprule
            \multicolumn{4}{l|}{SAC reference motivation}         & Categorization        &            &                &       &   &  \\ 
\midrule
         \rotatebox{45}{Explicit}  & \rotatebox{45}{Implicit}   & \rotatebox{45}{Beneficial} & \rotatebox{45}{Reference}          & \rotatebox{45}{Type}  & \rotatebox{45}{Market}         & \rotatebox{45}{Compliance} & \rotatebox{45}{Conformance} & \rotatebox{45}{Process}    & \rotatebox{45}{Product}      \\ 
\midrule
   \checkmark &            &             &  ISO/SAE 21434 DIS \cite{iso21434} & Standard       & International  &            & \checkmark  & \checkmark &  \checkmark  \\
   \checkmark &            &             &  SAE J3061 \cite{J3061}            & Guideline      & International  &            & \checkmark  & \checkmark & \checkmark   \\
              & \checkmark &             & SELF DRIVE Act \cite{SELFDRIVE}    & Regulation     & US     & \checkmark &             & \checkmark & \checkmark   \\
              & \checkmark &             & ADS 2.0 \cite{ADS2}                & Best-practise  & US     &            & \checkmark  & \checkmark & \checkmark   \\
              &            & \checkmark  & AV 3.0 \cite{AV3}                  & Best-practise  & US     &            & \checkmark  & \checkmark & \checkmark   \\
              &            & \checkmark  & GDPR \cite{gdpr}                   & Regulation     & Europe & \checkmark &             & \checkmark &  \\
              &            & \checkmark  & SPY Car Act \cite{SPYCar}          & Regulation     & US     & \checkmark &             & \checkmark & \checkmark \\
              &            & \checkmark  & CCPA \cite{CCPA}                   & Regulation     & US     & \checkmark &             & \checkmark &  \\
              &            & \checkmark  & UNECE GRVA CS \cite{UNECE-Reg}             & Regulation     &  International      & \checkmark &   & \checkmark & \checkmark   \\
              &            & \checkmark  & UNECE GRVA OTA \cite{UNECE-OTA}         & Regulation     &  International      & \checkmark &             & \checkmark & \checkmark \\ 
              &            & \checkmark  & ICV \cite{ICV}                     & Standard       & China  &           & \checkmark  &            & \checkmark   \\
              &            & \checkmark  & GB/T 35273 \cite{GBT35273}         & Standard       & China  &            & \checkmark  &        \checkmark    &    \\
              &            & \checkmark  & CSL \cite{CSL}                     & Regulation     & China  & \checkmark &             &        \checkmark    &    \\
\bottomrule
\end{tabular}
\end{table*}


There are both general and specific legislation and regulations regarding security and privacy that are applicable to the automotive industry. An example of the former is the European General Data Protection Regulation (GDPR) \cite{gdpr} and the Chinese equivalent GB/T 35273 \cite{GBT35273}, although a recommendation. Below we describe the relation between these documents and SAC. \\

Two of the analyzed documents explicitly mentions SAC: ISO/SAE 21434 DIS \cite{iso21434} and SAE J3061 \cite{J3061}. The former gives requirements on a cybersecurity case and that it shall provide the arguments that cybersecurity is achieved. The latter states that a cybersecurity case provides evidence and argumentation that design and implementation is sufficiently secure.

The Safely Ensuring Lives Future Deployment and Research In Vehicle Evolution (SELF DRIVE) Act~\cite{SELFDRIVE} contains requirements stating that a Cybersecurity plan that includes various security activities shall be developed. It thus implies that it shall be possible to show that the plan also has been implemented. The SAC shall therefor contain the corresponding evidence and argumentation of how the security plan has been met, including confidence in process for incident handling. 
Similarly, for the Automated Driving Systems (ADS 2.0) \cite{ADS2} it is beneficial to show evidence that cybersecurity processes are followed and show that the company adheres to industry best practice. The authors of the (ADS 2.0) document encourage documentation of how cybersecurity has been reached, including design choices, analyses, testing, etc. The SAC shall therefor cover these aspects.
The Automated Vehicles 3.0~\cite{AV3} also indicates the benefit to show evidence that cybersecurity processes are followed and show that the company adheres to industry best practice, including design principles and incident handling. The SAC shall therefor contain the corresponding evidence and argumentation of how security has been considered and handled during design and confidence in processes for incidence handling.

There is a current initiative on UN level, prepared by a subgroup of the working group on Intelligent Transport Systems / Automated Driving (IWG ITS/AD) of WP.29, referred to as ``UN Task Force on Cyber security and OTA issues'' (TF-CS/OTA), to establish regulations and type approval on Cybersecurity for vehicles that includes vehicle categories; M (standard passenger vehicles) and N (trucks). 
This include ``Regulation on uniform provisions concerning the approval of cyber security''~\cite{UNECE-Reg} and ``Regulation on uniform provisions concerning the approval of software update processes''~\cite{UNECE-OTA}.
For the former it can be seen as highly beneficial to show compliance and conformance to requirements in the regulation concerning process implementation and fulfillment as well as demonstration of performed verification activities and its outcome. The SAC shall therefor contain evidence and argumentation in regards to confidence in the implemented processes that shall cover the life-cycle of the vehicle to maintain an appropriate level of security. As well as product evidence and argumentation of an adequately secure product.
The latter one contain requirements for demonstration of evidence of a secure update process.
The SAC shall therefor contain evidence and argumentation in regards to confidence in the SW update processes in order to maintain an adequately secure product and acceptable level of risk.

Further, for the privacy related documents: General Data Protection Regulation (GDPR)~\cite{gdpr}; Security and Privacy in Your Car (SPY CAR) Act~\cite{SPYCar}; Chinese Information Security Technology – Personal Information Security Specification GB/T 35273~\cite{GBT35273}; and  California Consumer Privacy Act~\cite{CCPA}, SAC can be beneficial in regards to showing compliance to regulation, processes, methods and technologies to secure handling of data. The SAC shall therefor contain evidence and argumentation in regards to confidence in the processes as well as  implementation measures of handling private data. 

SAC can also be beneficial in the China Cyber Security Law (CSL)~\cite{CSL} to show that the organization has sufficient policies and processes that handle cybersecurity. Including incident handling and secure handling of private data.  The SAC shall therefor contain evidence and argumentation in regards to confidence in the post-production processes.
The CSL is a high level law and more detailed requirements will exist through the Chinese strategy and framework on Intelligent and Connected Vehicles (ICV)~\cite{ICV}. With the same reasoning it can be beneficial with the ICV in regards to showing compliance and conformance to requirements and processes, methods and design/technology. The SAC shall therefor contain evidence and argumentation of confidence in the implemented measures and processes to maintain an adequately secure product and acceptable level of risk.

%% file: results-RQ2.tex
\section{RQ2: Internal Needs and Oppotunities}\label{sec:rq2}

\subsection{Pre-study: Expectations of Security Leaders}
\label{sec:pre-study}

The panels of security experts concluded that the internal needs of an OEM in regards to the use of assurance cases for cybersecurity can be summarized as follows. 

\noindent\textbf{Secure product argumentation.} Need to have an argumentation for adequately secure implementations of HW/SW E/E systems in their vehicle throughout their life-cycle. This can be used, among other things, to make release decisions.

\noindent\textbf{Supporting evidence.} Need to have all the necessary evidence that the implemented systems are secure enough in order to release a vehicle to the end-customers. In particular, the evidence should undergo a quality assurance process and should cover all the relevant parts included in the development, with sufficient detail for the critical parts.
    
\noindent\textbf{Legal compliance.} Comply to the legislation, regulations and type approvals on national and international levels (in order to even be allowed to sell vehicles).

\noindent\textbf{Supply chain.} Manage collaboration between OEM and suppliers in terms of requirements, structure, aggregation and level of details for SAC constituents.

\noindent\textbf{Standard conformance.} Conform to standards, guidelines and best practices and implement state of the art processes and methods (in order to be competitive w.r.t. other OEMs).

\noindent\textbf{Process harmonization.} Existing development processes and the way of working need to be harmonized with the processes and methods required by SAC. One such example may be an agile way of working and product organization.

\subsection{Workshop to Identify Broad Usage Scenarios}
\label{sec:scenarios}

The workshop participants (working in three parallel group sessions) identified thirteen unique usage scenarios (US), which are listed below in no particular order. These scenarios depict, in a narrative forms, the broader set of needs and expectations of an OEM with respect to SAC. 
These scenarios cover a diverse set of stakeholders (10 different roles) within the OEM organization, which is, typically, a very large one.

\begin{figure}
    \centering
    \includegraphics[width=1\linewidth]{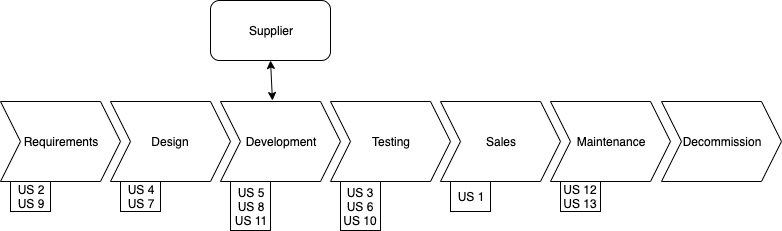}
    \caption{Usage scenarios for the life-cycle of an automotive product}
    \label{fig:lifecycle}
\end{figure}

When looking at the suggested usage scenarios, we can see that they span over multiple phases in an automotive product's life-cycle. Figure~\ref{fig:lifecycle} shows a high-level view of the different phases in the automotive product's life-cycle and the usage scenarios suggested in each phase. As the figure shows, the participant were able to identify at least one usage scenario in each phase except for the final decommission phase.

\noindent\textbf{US 1} As a salesman, I would use top-level SAC to prove to our customers that the company has considered all relevant security aspects of the final product, and has enough evidence to claim that it has fulfilled them.
 
\noindent\textbf{US 2}  As a member of the compliance team, I would use detailed SAC to prove to authorities that the company has complied to a certain standard, legislation, etc., and show them evidence of my claim of compliance.
 
\noindent\textbf{US 3}  As a project manager, I would use SAC to make sure that a project is ready from a security point of view to be closed and shipped to production.
 
\noindent\textbf{US 4}  As a project manager, I would include SAC in my project plan. I would make sure the project has the needed resources and time for creating the case (argumentation, evidence collection, etc.).
 
\noindent\textbf{US 5}  As a project would use SAC to monitor the progress of my project when it comes to fulfillment of security requirements.
 
\noindent\textbf{US 6}  As a product owner, I would use SAC to make an assessment of the quality of my product from a security perspective, and make a roadmap for future security development.
 
\noindent\textbf{US 7}  As a product owner, responsible for handling threats and vulnerabilities, I would use SAC to evaluate the effect of new threats and vulnerabilities, and evaluate whether a change is needed to the product.
 
\noindent\textbf{US 8}  As a member of the purchase team, I would include SAC as a part of the contracts made with suppliers, in order to have evidence of the fulfillment of security requirements at delivery time, and to track progress during development time.
 
\noindent\textbf{US 9}  As an action owner, I would use detailed and visual SAC to communicate with the risk owner, and decide how to update the product security in the right way (to know what to do)
 
\noindent\textbf{US 10}  As a system leader, I would use SAC to make an assessment of the quality of my system from a security perspective, and make a roadmap for future security development (same as US6, but on wider scope).
 
\noindent\textbf{US 11}  As a software developer, I would use SAC from previous similar projects as a guideline for  secure development practices.
 
\noindent\textbf{US 12}  As a legal risk owner, I would use SAC in court if a legal case is raised against the company for security related issues. I would use the SAC to prove that sufficient preventive actions were taken.
 
\noindent\textbf{US 13}  As a member of the corporate communication team, I would use SAC as a reference to answer security related questions.


\subsection{Prioritisation of Scenarios and In-depth Interviews}

\begin{figure}
	\centering
\begin{tikzpicture}
\begin{axis}[ 
xbar, xmin=0,
xlabel={Score},
symbolic y coords={
{US13},{US11},{US5},{US9},{US4},{US10},{US7},{US1},{US8},{US3},{US12},{US6},{US2}
},
ytick=data,
nodes near coords, 
nodes near coords align={horizontal},
ytick=data,
]
\addplot[fill=gray] coordinates { 
(33,{US2}) 
(23,{US6})
(21,{US12})
(19,{US3})
(18,{US8})
(11,{US1})
(9,{US7})
(6,{US10})
(5,{US4})
(4,{US9})
(1,{US5}) 
(0,{US11}) 
(0,{US13}) 
 };
\end{axis}
\end{tikzpicture}
\caption{Prioritized usage scenarios} \label{fig:top5}
\end{figure}
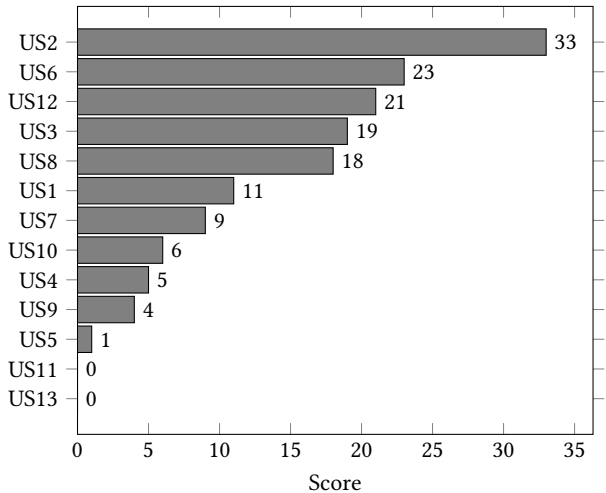

As mentioned in the methodology, after the workshop we sent the scenarios to experts and asked them to prioritize them based on the value the scenarios provide to the company. The result of the prioritization task is shown in Table \ref{fig:top5}. For the top five scenarios, we identified the key stakeholders for those scenarios and conducted in-depth interviews.

\paragraph{Interview on US3 and US6 -- Product delivery and process improvement}
The interview was conducted with two interviewees, to tackle two of the usage scenarios, as explained in the methodology~\ref{subsec:methodology}.
The interviewees were one solution train engineer and one solution manager (participants 19 and 20 in Table \ref{tbl:participants}).
\begin{aquote}{Participant 19}
[SAC provides] an opportunity for the company to build a reputation of building secure vehicles.
\end{aquote}
The interviewees said that today, the security assurance in projects is mostly based on experience, and is done by providing evidence such as test reports for some claims. These claims are derived from projects' requirements are not put together in a structured way. The interviewees also stressed that what is done today is simply not sufficient given the rapid evolution of connectivity in the vehicles. Historically there has been \emph{``a trust in the physical shell (the cab) around the items in the vehicle.''}, but today these items are connected to the outside world, and can be a target for cyber attacks.
 
As per the interviewees, having SACs as a part of scoping projects, setting the milestones and deliverables, and connecting them to the development process, would be of great value to security assurance from a project management perspective. Additionally from a product point of view, having a holistic SAC which is updated by the related projects would give the product owner an understanding of how secure the product is at different time-points, e.g., after integrating changes from different projects.

Both interviewees agreed that SACs \textbf{should be built on a product level rather than a project level}. SACs should be integrated and built within projects, but only to contribute towards the product's SAC. A product, however, may be big and complex, and may include items that are more interesting from a security perspective than others, e.g., Electrical Control Units (ECU) that are connected to the outside world in a vehicle. In this case there needs to be a severity assessment to focus on the right items.

The interviewees think that the SAC work can be integrated into the project manager's and product owner's work by capturing it in projects' requirements. They emphasized that it is important to work on quality, e.g., security as a \textbf{part of the development process, and not as a separate activity}. The workload in the beginning will be high, but with time, there will be patterns that can be reused to build SAC with less effort.

When it comes to the challenges, the interviewees consider handling the complexity of the product (vehicle in this case), and finding right competences to be the main ones. Other challenges include securing a buy-in to work with SAC all the way from upper management to development teams.

\paragraph{Interview on US2 -- Compliance} 
The interview was conducted with a functional safety assessor who has a wide range of experience in compliance (participant 21 in Table~\ref{tbl:participants}).
\begin{aquote}{Participant 21}
The security case can serve as an umbrella document for all the analysis documents.
It can be used by an assessor to find the right documents in order to assess compliance to a standard.
\end{aquote}
In the current situation, the compliance team is only concerned with safety matters, as per the interviewee. The only involvement with security issues is when there is a breach which affects the product's safety. The expectation is that SAC can \textbf{serve as an umbrella document for all the analysis documents}. Hence it can be used by an assessor to find related document to assess compliance to a certain standard.

The SAC should be \textbf{created on a whole vehicle level}, as per the interviewee. It needs, however, to be \textbf{dividable in manageable pieces}, i.e., pieces that can be managed by individuals, which would be responsible for the corresponding part of the SAC. For compliance purposes, the authorities look at system level, hence, for every project that includes changes in the system, it is important to conduct an \textbf{impact analysis} to identify affected artefacts, and update the SAC accordingly. 

How working with SAC can be integrated within the way of working of the compliance team depends much on how SACs are implemented. If they are integrated within the projects (as they should according to the interviewee), then the compliance team would have the responsibility of following them up throughout the project, as well as making sure that they are complete after the verification phase. 

The interviewee considers the main challenge to be finding resources and competences to carry out the work related to creating, maintaining, assessing and supporting SACs.

\paragraph{Interview on US8 -- Suppliers}
The interview was conducted with a component owner in Company B, who is experienced in purchasing and working with suppliers (ID 22 in Table \ref{tbl:participants}).
\begin{aquote}{Participant 22}
[Working with SAC provides] An opportunity to catch up with suppliers, which in many cases have come further in thinking about security than the company.
\end{aquote}
Today, security assurance when working with suppliers is about making sure of the fulfillment of security requirement, and running test cases on a sample of the requirements. However, \emph{``there is an uncertainty to a large extent that the received software is secure''}, as per the interviewee. On the other hand, safety critical functions, e.g., breaking, is handled differently. The suppliers are usually asked to show how requirements are broken down and implemented during regular review sessions. In some cases, suppliers are asked to provide safety cases, which are used together with the internal safety cases to make sure that the claims align. The interviewee expects that SAC can be used in the same manner as safety cases. Additionally, they can be used to \textbf{communicate} regarding security both with suppliers and internally, and as a \textbf{supporting artifact for creating security requirements}.

Regarding the granularity of the cases, the interviewee distinguishes between two types of SAC: the ones created at the supplier's end, and the ones created by the company.
On the supplier's side the cases should be on the ECU level, followed by a threat-based level. Whereas the company-owned SACs should be on the complete vehicle level, and broken down to ECU level, which is contributed by the suppliers. An important aspect mentioned by the interviewee is the weighting of the claims based on severity. This is to be able to prioritize the claims during the follow up sessions with the suppliers, and when testing the implementation.

Integrating SAC in the current way of working with suppliers would increase the workload, but there are no obvious conflicts, according to the interviewee. However, there is a \textbf{need for tools} to store, extract, and compare SACs. Additionally, a version handling tool is also required to keep track of the SACs and their changes. The interviewee also mentioned the need to \textbf{use an exchangeable format} when building the cases, on both ends (supplier and company), in order to compare and integrate them.

A challenge is to find and provide practical training on SAC in the industrial context. Another challenge mentioned by the interviewee is finding resources with the right competences to carry out the SAC related work. The interviewee emphasized that based on the experience from safety, even when there is education about the cases, it was much more complicated when actual work was done.

\paragraph{Interview on US12 -- Legal}
The interview was conducted with a senior legal counsel at Company B (participant 23 in Table~\ref{tbl:participants}). 
\begin{aquote}{Participant 23}
An evidence based structured approach to argue about security would definitely be used as evidence in court.
\end{aquote}
In the current situation, the company has not had any legal case for security related issues, but there have been functional-safety related cases claiming that feature malfunctioning have caused accidents. 
Current evidence used in court for these kinds of legal cases are of two types: \emph{(i)} usage of technology according to an acceptable standard; \emph{(ii)} implementations of the used standards and technologies are correctly done (this should be certified by a third party assessor). Hence, if the ISO 21434 \cite{iso21434} becomes an industry standards, then it can be used as an evidence in court.  However, it is very important to assure the quality of the case when it comes to completeness in the argumentation and evidence. The SAC used as an evidence will be available to the opponents, and it could be exploited to find holes and error to be used against the company.

The granularity needed for the SAC depends pretty much on the legal case according to the interviewee. However, there is a need to create \textbf{SACs for complete vehicles, and views that could be broken down}. This is to avoid cases where a legal case against the company involves a composition of systems (end-to-end function), and SACs are created for a subset of these systems. This would be a weakness if they are used as evidence.

The creation of SACs should, from a company perspective, be \textbf{during development} to assure security, according to the interviewee. However, from a legal perspective, SACs can be created once a legal case is filed, but that could lead to questions asking why the SACs were not created during development. Additionally, it would increase the probability of the risk of having insufficient evidence, as it would be harder to locate and assign them. Moreover, from a liability perspective, it is much better to create the SAC proactively.

As per the interviewee, if SACs are to be used as evidence, the relevant stakeholders in the company will be reached out and asked to provide the SACs when needed. Then the legal responsible will have meetings with the stakeholders to understand it. This means that the ownership of the SACs would not be the responsibility of the legal responsible, even if it is created specially for a legal case. The legal responsible would be a user of the SAC.

At the end, the interviewee stressed that introducing a structured way of security assurance \emph{''can lead to creating better systems which can protect the company from issues from regulators and third parties.''}. However, there has to be a buy-in on different levels in the company in order to do this in a correct way.

%% file: threats.tex
\section{Threats to validity}
\label{sec:threats}

In terms of \emph{external validity}, we are aware that the general validity of our results could be limited to the companies involved in the study. Also, the companies are from the same country. Therefore, the results might not directly translate to companies with a different culture.
However, the involved companies are of high profile, quite large and compete at the international level. Therefore, they are able to provide a quite broad perspective on the entire automotive industry.
In any case, the results presented in this paper are an important first important step towards a larger survey study involving more companies and professionals, internationally.

In terms of \emph{internal validity} we consider the following aspects. First, the selection of the standards and regulations investigated in RQ1 could have been incomplete. However, we are confident we are adressing the most important documents, especially for the mentioned markets (EU, China, US). 
Second, in the prioritization of the scenarios of RQ2, there is a risk that the selection of the top scenarios was biased by present market pressure towards compliance to the upcoming standards.
Third, elements of bias could be have been introduced via the selection of the participants. 
The same limitation applies to the participants of the interviews, as the selection of participants was based on expertise and availability (convenience sampling). 
All in all, we have a balance mix of participants with different types of expertise: security, product development, business, and legal. This provides us with enough confidence that the results are representative of the expectations and needs across the studied companies. 

%% file: discussion.tex
\section{Discussion}
\label{sec:discussion}

As this study contains different parts, we think it might be useful to the reader to if we illustrated how the different results relate to each other. 
Further, we translate the results into a concrete set of recommendations for the OEMs.

\subsection{Mapping of Results}

\begin{table*}
\caption{Traceability of the results in this study}
\label{tbl:coverage}
\begin{tabular}{lll}
\toprule
         Internal needs: pre-study & Internal needs: Usage scenario & External needs \\ 
\midrule
  Secure product argumentation & US3, US5, US6, US7, US9, US10, US11 
                               & \cite{iso21434},
                                 \cite{J3061}    \\
  Supporting evidence          & US2, US8, US11, US12 
                               & \cite{iso21434},
                                 \cite{J3061}  \\
  Legal compliance             & US2, US8 
                               & \cite{gdpr},
                                 \cite{SELFDRIVE}, 
                                 \cite{SPYCar}, 
                                 \cite{CSL},
                                 \cite{CCPA}, 
                                 \cite{UNECE-Reg},  
                                 \cite{UNECE-OTA}    \\
  Supply chain                 & US2, US3, US5, US8 
                               & \cite{gdpr},
                                 \cite{iso21434},
                                 \cite{ADS2},
                                 \cite{J3061}, 
                                 \cite{GBT35273}, 
                                 \cite{UNECE-Reg},
                                 \cite{UNECE-OTA},
                                 \cite{AV3}   \\
  Standard conformance         & US6, US8, US10, US11 
                               & \cite{ICV},
                                 \cite{iso21434},
                                 \cite{ADS2},
                                 \cite{J3061},
                                 \cite{GBT35273},
                                 \cite{AV3} \\
  Process harmonization        & US4, US5  &     \\
  ---						   & US1, US13 & \\
\bottomrule
\end{tabular}
\end{table*}

Table~\ref{tbl:coverage} shows a mapping of the results we got from the different parts of this study: the internal needs from the pre-study, the usage scenarios, and the external needs. 
As an example of the mapping, using SAC as \emph{Supporting evidence} is identified as an internal need in the pre-study. It relates to multiple usage scenarios, e.g., \emph{US12}, where SAC can be used as a piece of evidence in case of a legal suit. It also relates to the ISO/SAE 21434 DIS~\cite{iso21434} standard and the SAE J3061~\cite{J3061} best practice which explicitly require SAC with evidence on secure design and implementation.

As shown in the table, the results are not heterogeneous, but rather align to a large extent. 
Every internal needs identified by the experts in the pre-study is linked to at least one of the usage scenarios, as well as one external driver, with the exception of the \emph{Process harmonization} need.
The table shows how the scenarios proved to be useful tool to obtain a deeper view into the internal needs and in detailing the high-level needs identified in the pre-study.
On the other hand, the table also show that the scenarios go beyond the high-level needs identified by the security leaders, by broadening the scope of the analysis.
In particular, two additional scenarios (\emph{US1} and \emph{US13}) suggest using an abstracted level of SAC to communicate and answer security related questions both from inside the company and from potential customers.

\subsection{Recommendations}

As a summary of the findings of this study, we provide our recommendations for companies (particularly, OEMs) that are starting to work with security assurance cases. 
These recommendations are not complete solutions, but rather steps towards establishing a ground to integrate SAC with the companies' way of working, and to help different stakeholders make use of SAC.

\emph{Standards and Regulations -- Cover both process and product.}
Several of the security-related standards/regulations contain both requirements on processes and the product. The processes include how to develop the product in a secure manner as well as keeping the product secure after its release. In some cases the product requirements even suggests what kind of measure that should be considered. Since both standards and regulations put requirements on audits and assessments of processes and products, including certification of security processes, the SAC can function as the tool for showing both compliance and conformance to those requirements.
Privacy related standards/regulations foremost imply that processes shall be in place in order to handle private data in a secure manner. This also means having the appropriate measure in place in order to accomplish this. The conclusion is thus that the SAC shall contain arguments and evidence both that the processes are sufficient and that the implementations of mechanisms that handle private data are secure. Some of the regulations mentioned in \autoref{tbl:external} that foremost requires processes for handling of data, such as \cite{gdpr},  \cite{CCPA} and \cite{GBT35273} can implicitly be considered as product requirements since there will also be a need for implementation of security mechanisms that for example encrypt data during transfer. 

\emph{Granularity of SAC -- Whole product over sub-projects.} 
In industries producing complex products, e.g., automotive, it is common that the products are organized in multiple projects. Additionally, the changes to these products are also done using projects (commonly called delta projects). In this case, SACs should be created on a product level rather than a project level to fulfill the usage scenarios identified in our study. When a SAC is created within a project, it has to be integrated later with the product's SAC. They should, however, be built in a way which allows different stakeholders to have different views corresponding to different abstraction levels.

\emph{Creation of SAC -- No retro-fitting and no silos.} 
It is possible to build SAC for existing products, but going forward, it is important to embed the work on SAC into the development process at the organization.
Security cases can provide real value to an organization if it is not just considered as a ``check-in-the-box'' activity, or, worse, an overhead.
Security cases can be used as a methodology for guiding the work of cybersecurity and communicating its importance across the teams.
The work on Security cases should include a large part of the organization, with different teams working on different parts of the security case, e.g., teams working on the system, component, and functional levels. 
This requires to have clear collaboration interfaces to avoid issues such as inconsistencies, and conflicts. 
Moreover, each case should have an owner (even SACs that consist of multiple sub-cases), which should be made explicit to all teams working on the cases. This owner would drive the work on the SAC and be responsible for its maintenance and quality control.
However, in order for this to work, there must be a buy-in on different level of the organization, starting from top management, and continuing all the way through the organization to the development teams. Additionally, the importance of tool support and automation should not be underestimated.


\emph{Quality of SAC -- Actively assess completeness and confidence.} 
Security assurance cases are going to serve multiple purposes within the organization (see the diversity in the usage scenarios) and they can be used in contexts that have different levels of criticality (e.g., legal case vs process improvement). Therefore, it must be clear what the quality level of each SAC is, so that they are not used in the wrong context.  
We recommend to introduce measures to assess the quality of the SACs. These may include, for example, the \emph{completeness} of the argumentation, and the level of \emph{confidence} in the evidence. This is emphasized by the security experts in the pre-study (second bullet in Section \ref{sec:pre-study}), as well as all the interviewees. However, in some of our identified usage scenarios, this becomes even more important. For example when SAC are used as evidence in court, they would be available to the oppositions as well, meaning that any flaw could back-fire against the company. Hence, it is very important to be able to assess the quality of a SAC before using it in court.

\emph{SAC and suppliers -- A common language is key to smooth collaboration.} When it comes to working with suppliers, the SAC should be built using an exchangeable format. This is to enable the SAC created by the suppliers to be integrated with the SAC of the corresponding product. Another important aspect is to add an assessment of the severity level of the claims. This is to be able to followup on the most severe items, when it is not possible to do a complete followup. 

\emph{SAC and suppliers -- Plan for shared ownership.}
The suppliers might require to keep parts of the SAC private (e.g., some evidence). In this case, it is important to have a mechanism to keep ensuring the overall quality of the SAC, e.g., by introducing a black-box with meta-information such as the validity of the items behind the box. Additionally, the ownership of the whole case has to be considered, as the complete SAC would not be in the hands of a single stakeholder. For instance, using a SAC in a legal case would require a disclosure process in order to compile a SAC from multiple sources and multiple owners.

\emph{Miscellanea -- With opportunities come challenges.}
Working with SAC is not trivial and comes with many challenges. Traceability and change analysis were considered main challenges by the majority of the participants. Additionally, finding the right competences to carry out the SAC-related work, role identification and description, and acquiring the right tools and integrating them in the organizations tool chain were also considered major challenges.


%% file: conclusion.tex
\section{Conclusion}\label{sec:end}


In this study, we have analyzed the requirements around the use of security assurance cases in the automotive domain. In particular, we have listed the constraints coming from standards \& regulations and have identified the internal needs of OEMs.
We have concluded this work by translating the results in a number of pragmatic recommendations for OEMs.
These are valuable and necessary contributions for the overall furthering of the automotive industry and for a more effective and secure development of future road vehicles.

In future work, we plan on extending this work with a survey including a larger group of automotive companies and automotive security experts.
Additionally, we are addressing the requirements identified in this paper by means of a systematic methodology to create security assurance cases for the automotive industry.
